\documentclass[iop]{emulateapj}
\usepackage{amsmath, amssymb}
\usepackage{amsfonts}
\newcommand   *{\B}[1]       {\boldsymbol{#1}}
\newcommand   *{\Msun}       {{\mathrm{M}_{\odot}}}
\renewcommand *{\d}          {\mathrm{d}}
\newcommand   *{\peri}[1]    {#1_{\scriptscriptstyle -}}
\newcommand   *{\apo}[1]     {#1_{\scriptscriptstyle +}}
\def\msun{{\rm M_{\odot}}}

\def\be{{\begin{equation}}}
\def\ee{{\end{equation}}}

\begin{document}

\title{X--ray Transients: Hyper-- or Hypo--Luminous?}
\author{Jean--Pierre Lasota\altaffilmark{1,2,3}, Andrew R. King\altaffilmark{4,5} and Guillaume Dubus\altaffilmark{1,6,7}}
\altaffiltext{1}{CNRS, UMR 7095, Institut d'Astrophysique de Paris, 98bis Bd Arago, 75014 Paris, France; lasota@iap.fr}
\altaffiltext{2}{Nicolaus Copernicus Astronomical Center, ul. Bartycka 18, 00-716 Warszawa, Poland}
\altaffiltext{3}{Sorbonne Universit\'es, UPMC Univ Paris 06, UMR 7095, 98bis Bd Arago, 75014 Paris, France}
\altaffiltext{4} {Theoretical Astrophysics Group, University of Leicester, Leicester LE1 7RH, U.K.; ark@astro.le.ac.uk}
\altaffiltext{5}{Astronomical Institute Anton Pannekoek, University of Amsterdam, Science Park 904, 1098 XH Amsterdam, Netherlands}
\altaffiltext{6}{CNRS, IPAG, F-38000 Grenoble, France}
\altaffiltext{7}{Univ. Grenoble Alpes, IPAG, F-38000 Grenoble, France, Guillaume.Dubus@obs.ujf-grenoble.fr}

\begin{abstract}
The disk instability picture gives a plausible explanation for the behavior of soft X--ray transient systems if self--irradiation of the disk is included. We show that there is a simple relation between the peak luminosity (at the start of an outburst) and the decay timescale. We use this relation to place constraints on systems assumed to undergo disk instabilities. The observable X--ray populations of elliptical galaxies must largely consist of long--lived transients, as deduced on different grounds by \citet{PiroBildsten02}.  The strongly--varying X--ray source HLX--1 in the galaxy ESO 243--49 can be modeled as a disk instability of a highly super--Eddington stellar--mass binary similar to SS433. A fit to the disk instability picture is not possible for an intermediate--mass black hole model for HLX--1. Other, recently identified, super-Eddington ULXs might be subject to disk instability.
\end{abstract}

\keywords{black hole physics --- accretion, accretion disks -- X--rays: binaries}

\section{Introduction}
\label{sec:intro}

Many bright X--ray sources are strongly variable. It is now largely accepted that much of this variability results from the thermal--viscous disk instability \citep[see][for a review]{Lasota01}. The instability -- originally discovered for dwarf novae, which are accreting white dwarf systems -- results from the presence of ionization zones of hydrogen in the accretion disk (helium in some some ultracompact systems). The disk is forced to alternate between quiescence, when the disk is cool and faint and hydrogen predominantly neutral, and outbursts in which the disk is hot and hydrogen ionized. If the accretor is a neutron star or black hole, the X--rays produced by central accretion keep the disk in the hot state and only allow a return to quiescence on a viscous timescale. 

Although some of the properties of this model remain to be worked out, in particular the duration of the quiescent phase, its predictions for outburst behaviour are robust enough to allow quantitative tests \citep{Coriat12}. We show here that there is a simple connection between the accretor mass, the peak luminosity at outburst and the decay time of the outburst, if this is well--defined by observations. If central irradiation is able to keep the whole disk in the hot state the outburst indeed has a fast--rise, exponential decay (FRED) shape \citep{KingRitter98,Dubus01}.

The connection we consider is particularly useful for uncovering the properties of varying X--ray sources which are too distant, or whose distance is too uncertain, to study easily in other ways. Here we apply it the the X--ray populations of elliptical galaxies, and to the strongly--varying X--ray source HLX--1, whose nature remains uncertain.

\section{Observable properties of disk instabilities}
\label{sec:obsdim}

Two observable properties characterize the disk instability picture for X--ray transients. First, the maximum accretion rate (at the start of the outburst) is that of a (quasi)steady X-ray irradiated disk accreting at constant rate of $\sim 3\dot{M}_{\rm crit}^{+}(R_d)$, where $\dot{M}_{\rm crit}^{+}(R_d)$ is the value of the minimum critical accretion for a hot, irradiated disk at its outer radius $R_d$ \citep[see e.g. Fig. 31 in][]{Lasota01}.

This fixes the critical accretion rate $\dot M_{\rm max}$ through the relation \citep{Lasota08}
\begin{eqnarray}
\dot M_{\rm max}\approx 3 \dot{M}_{\rm crit}^{+} &=&7.2 \times 10^{17}
                     ~{\mathcal  C}_{-3}^{-0.36}
                      ~\alpha_{0.2}^{ 0.04+ 0.01\log{\mathcal C}_{-3}}\nonumber \\
                    &&R_{d,11}^{2.39-0.10\log{\mathcal C}_{-3}}
                      ~m^{-0.64+ 0.08\log{\mathcal C}_{-3}}\,\rm g\,s^{-1}\nonumber \\
\label{temp1}
\end{eqnarray}
where ${\mathcal  C}=10^{-3}{\mathcal  C}_{-3}$ is a constant characterizing the outer-disk irradiation by the point-like source centered at they accretor \citep{Dubus99}, $m$ is the accretor mass in solar units, $\alpha < s1$ the standard viscosity parameter, and $R_{d}=R_{d,11}10^{11}$cm is the disk outer radius.
Taking ${\mathcal  C}_{-3}=~1$ \citep{Dubus01} and ignoring the very weak dependence on $\alpha$, one gets
\begin{equation}
\dot{M}_{\rm max} \approx 7.0 \times 10^{17}~R_{d,11}^{2.39}~m^{-0.64}\,\rm g\,s^{-1}.
\label{temp}
\end{equation}
FRED--type X--ray transients  typically have disk radii $\sim 10^{11}$cm, corresponding to orbital periods of $\sim 10$ hr.

The second observable property is the decay time for the X--rays: in soft X--ray transients (unlike dwarf novae), disk irradiation by the central X--rays traps the disk in the hot, high state, and only allows a decay of $\dot M$ on the hot--state viscous timescale \citep{KingRitter98,Dubus01}.
This is
\begin{equation}
t \simeq \frac{R^2}{3\nu}
\label{t}
\end{equation}
Here the Shakura--Sunyaev viscosity is $\nu = \alpha c_s^2/\Omega$, where $c_s \propto T_c^{1/2}$ is the sound speed, $T_c$ the disk midplane temperature, and
$\Omega = (GM/R^3)^{1/2}$.
This gives
\begin{equation}
t \simeq \frac{(GMR)^{1/2}}{3\alpha c_s^2}.
\label{tvisc1}
\end{equation}
Taking the critical midplane temperature $T_{\rm c}^{+}\approx16300\,\rm K$ corresponding to $\dot{M}_{\rm crit}^{+}$ \citep{Lasota08}\footnote{For irradiated disks, the value of $T_c$ depends
only weakly on viscosity and irradiation and is independent of mass and radius.} one obtains
for the decay timescale
\begin{equation}
t\approx 43   \ m^{1/2} R_{d,11}^{1/2} \alpha_{0.2}^{-1} \,\rm days,
\label{tvisc}
\end{equation} 
where $\alpha_{0.2}=\alpha/0.2$.  In the thermal-viscous disk instability model (hereafter TVDIM) the critical temperatures
depend only on the ionization state of the disk matter  and are thus practically independent of radius, viscosity parameter etc. \citep{Lasota01,Lasota08}.
By definition, the viscous decay time depends on the parameter $\alpha$. From observations of dwarf nova outbursts one deduces that
$\alpha \approx 0.2$ \citep[see][]{Smak99,KP12}. Comparison of models with observations of X-ray transients suggests the same value of $\alpha$ for these
systems too \citep[][see also Sec. \ref{sec:subedd}]{Dubus01,KPL07}.

Eliminating $R$ between (\ref{temp}) and (\ref{tvisc}) gives the
peak accretion rate through the disk at the start of the outburst as
\begin{equation}
\dot{M}=4.9 \times 10^{17}\ m^{-3.03} \left(t_{40}\alpha_{0.2}\right)^{4.78} \rm g\,s^{-1},
\label{mdot}
\end{equation} 
with $t = 40\, t_{40}\,{\rm d}$. Assuming an efficiency of $\eta$ of 10\%, the corresponding  luminosity  is
\begin{equation}
L=4.4 \times 10^{37}\ \eta_{0.1} m^{-3.03} \left(t_{40}\alpha_{0.2}\right)^{4.78} \rm erg\,s^{-1}
\label{luminosity1}.
\end{equation}

 \section{Sub--Eddington Outbursts}
 \label{sec:subedd}

The observed behaviour of outbursting systems differs significantly depending on whether they have sub-- or super--Eddington accretion rates. The Eddington accretion rate is
\begin{equation}
\dot M_E = 1.3\times 10^{18}\eta_{0.1}^{-1}m~{\rm g\, s^{-1}}
\label{mdotedd}
\end{equation}
where $\eta = 0.1\eta_{0.1}$ is the accretion efficiency
we find the Eddington accretion ratio
\begin{equation}
\dot m = 0.34 \eta_{0.1}(\alpha_{0.2}t_{40})^{4.78}m^{-4.03}.
\label{eddratio}
\end{equation}

This  equation shows that the start of the outburst is sub--Eddington only if the outburst decay time is relatively short or the accretor (black hole) mass is high, i.e. the observed decay timescale is
\begin{equation}
 t \la 50\, \eta_{0.1}^{-0.21}\alpha_{0.2}^{-1}m^{0.84}~{\rm d},
\label{subedd}
\end{equation}
in good agreement with the decay timescale of the detailed outburst models of \citet{Dubus01} and, more importantly, with the compilation of X--ray transients outburst durations by \citet{YanYu14}. This suggests that  the standard value of  $\eta_{0.1}\simeq1$,  and the value $\alpha_{0.2}\simeq1$ deduced from observations of dwarf novae, give the correct order of magnitude for the decay timescale in this type of system (from $\approx 3$ days to $\approx 300$ days, Fig. 5 in \citealt{YanYu14}). This equation also implies that black hole transients have longer decay timescales than neutron star transients, all else being equal. Indeed, \citet{YanYu14} find outbursts last on average $\approx 2.5\times$ longer in  black hole transients than in neutron star transients. 
For sub--Eddington outbursts Eq.~(\ref{luminosity1}) provides a straightforward relation between luminosity and outburst decay time. For a decay timescale $t$ of 0.5 years (see Sec. \ref{sec:appsub}), the expected luminosity is 
\begin{equation}
L = 6.2 \times 10^{40}\eta_{0.1}(\alpha_{0.2}t_{0.5\,\rm yr})^{4.78}m^{-3.03}~{\rm erg\, s^{-1}}
\label{lt}
\end{equation}
Equivalently, this gives a relationship between distance $D$, bolometric flux $F$ and outburst decay time $t$, 
\begin{equation}
D_{\rm Mpc} \simeq 23\, m^{-1.5}\left(\frac{\eta_{0.1}}{ F_{12}}\right)^{1/2}(\alpha_{0.2}t_{0.5\,\rm yr})^{2.4}
\label{flux}
\end{equation}
where $D = D_{\rm Mpc}\,{\rm Mpc}$ and  $F = 10^{-12}F_{12}\,{\rm erg\, s^{-1}\,cm^{-2}}$.

\section{Super--Eddington Outbursts}
\label{sec:superedd}

If the condition (\ref{subedd}) does not hold, the initial outburst accretion rate is super--Eddington.
This has two consequences. First, because the accretion luminosity is at the radiation pressure maximum for a range of radii the bolometric luminosity is larger than the Eddington limit by a factor $\sim 1 + \ln\,\dot m$ \citep{SS73,Poutanen07}. Second, the outburst luminosity is likely to be beamed, and we adopt here the form deduced by \citet{King09}.  
When $\dot{m}\ga 8$, an observer situated in the beam of this outbursting system (ULX) infers an apparent (spherical) X--ray luminosity
\begin{equation}
L_{\rm sph} =\frac{1}{b}L_E[1+\ln(1 +\dot m)]
\end{equation}
with the beaming factor $b$
\begin{equation}
b \simeq \frac{73}{\dot m^2},
\label{beam}
\end{equation}
i.e. 
\begin{equation}
L_{\rm sph}= 2.2\times 10^{36}\dot m^2m[1+\ln(1 +\dot m)]~{\rm erg\,s^{-1}}
\label{lsph}
\end{equation}
 \citep{King09}. Since the apparent luminosity scales approximately as $\dot m^2$, and $dL/dt=(dL/d\dot M)(d\dot M/dt)$, the observed decay time is now 
\begin{equation}
t = \frac{R^2}{6\nu} \label{tbeam}
\end{equation}
leading to the modified form of (\ref{eddratio}) as
\begin{equation}
\dot m = 1.3\times 10^4 \eta_{0.1}(\alpha_{0.2}t_{0.5\,\rm yr})^{4.78}m^{-4.03}.
\label{eddratio2}
\end{equation}

The X--ray light curve is now determined purely by beaming, through Eq. (\ref{lsph}).
The outburst appears to decay twice as fast, as the beaming reduces simultaneously with the accretion rate. Equations (\ref{lsph}, \ref{eddratio2}) determine the accretor's mass $m$
and the Eddington ratio $\dot m$ for any given case.

\section{Application to Observed Sub--Eddington Systems}
\label{sec:appsub}

We first use equations (\ref{lt}, \ref{flux}) for extragalactic binary systems. We see that these are unlikely to be detectable if the outburst decay is short ($\la $ years), particularly if the accretor is a stellar--mass black hole ($m \ga 3$). This independently supports the conclusion of \citet{PiroBildsten02} that the X--ray populations of elliptical galaxies probably consist largely of soft X--ray transients undergoing outbursts so long that that the ensemble shows little observed variation. 
They reached this conclusion since if the X--ray population were genuinely persistent, the observed accretion rates would long ago have exhausted the masses of the (necessarily low--mass) donor stars and extinguished the X--rays. If instead these systems are in reality transient, it follows that they must have very short duty cycles $d$. The total population must be larger by the factor $1/d \gg 1$, but almost all these systems are in quiescence at any given epoch. The required long outbursts tell us that the systems must have large disks, and so are wide, implying evolved donor stars. For such long periods the outburst morphology is too complex for the simple equations of Section 3 to hold. Accordingly we cannot use eqn (\ref{lt}) to argue from the fact that 
the known sources often have modest luminosities that they must in general have black hole accretors ($m \ga 3$).

Equations (\ref{lt}, \ref{flux}) also show that at extragalactic distances, sub--Eddington transients with black hole masses above the normal stellar range (i.e. intermediate--mass or supermassive)
must either be very faint, or have extremely long decay times. In other words such systems cannot correspond to observed rapidly--decaying transients unless these are within the Galaxy. 

This argument is relevant for the source HLX--1 \citep{Farrell09}. This shows a sequence of quasi--regular outbursts lasting $ < 180$ days. For the four outbursts between 2009 and 2012 the recurrence time was $\sim 370$ days, but the 2013 outburst started about 1 month `late' \citep{Godet13a,Godet14}. Multiwavelength observations of HLX--1 during 2009--2013 reveal outburst properties similar in most respects to those of low--mass X--ray binaries \citep[LMXBs: e.g.][]{RMcC06}. HLX--1 is positionally coincident with
the outer regions of the edge--on spiral galaxy ESO 243--49 \citep{Farrell09} at a redshift of 0.0224.\footnote{A narrow $H_\alpha$--line observed with the same redshift \citep{Wiersema10,Soria13} suggests that the positional coincidence corresponds to real membership.} At the distance ($D = 95$~Mpc) of this galaxy its unabsorbed 0.2--10~keV luminosity $L_{\rm max} = 1.3\times 10^{42}$~erg\,s$^{-1}$ at maximum requires a black hole mass 
 $m> 10^4$, for the source to be sub-Eddington radiator. 
But from Eq. (\ref{lt}) this requires a decay time of order decades or centuries, in complete contrast to the observed $\lesssim$ 180 days. This supports the conclusion of \citet{Lasota11}, who found that a disk instability model could not explain the light curve of HLX--1 if the system was assumed to contain an intermediate-mass black hole (IMBH) and be at the 95 Mpc.

According to Eq. (\ref{eddratio}), an X-ray transient with outburst duration of $\sim 180$ days is sub-Eddington if the black-hole mass $m \ga 4.5$. This in turn corresponds to a distance $< 2.3$ Mpc. 
For $D \sim 1$ Mpc, say, we have $m\simeq 8$ and $R\sim 2.2\times 10^{11}$\, cm and maximum outburst luminosities $L_{\rm max} \sim 1.1 \times 10^{38}{\rm\, erg\, s}^{-1}$.  Such a system would have a period $\ga 1$ day, so probably has an evolved donor star. 
Many low--mass X--ray binaries (LMXBs) like this are known, and almost all are transient. 
For HLX--1 to be a standard bright transient of this type the required distance 
 $D \sim 1$\, Mpc places it outside the Milky Way but still well within the Local Group. The system must reside in a nearby dwarf galaxy. These are known to harbor transient LMXBs \citep{Maccarone05}. There are probably far more dwarf galaxies in the Local Group than so far discovered \citep[see e.g.][]{Koposov08}. Because the escape velocities from them are far smaller than the typical natal kick velocities of LMXBs 
 these can either escape them altogether or have orbits which put them well outside any visible host for most of their lives \citep{DehnenKing06}.  These considerations mean that HLX--1 could turn out to be hosted by a previously unrecognized Local Group dwarf galaxy, but might also have no apparent host. Were its association with ESO 243--49 to be challenged, HLX--1 could well be such a system. 
In summary, at least one of the three statements
 
 \noindent
 (a) HLX--1 is at 95 Mpc
  
 \noindent
 (b) its outbursts are driven by the thermal--viscous disc instability
 
 \noindent
 (c) it is sub--Eddington
 
 \noindent
 cannot be correct.

 Adopting (a), we must drop at least one of (b) or (c). If this is (b) we must consider models where
 the outbursts of HLX--1 differ fundamentally from those of all other soft X-ray transients. 
 This appears inherently unlikely, and worse, there is little room for plausible alternatives. In sub--Eddington stellar--mass binaries the only other outburst model seriously considered (e.g. for Be X--ray binaries)  invokes periodic mass transfer from a companion on an eccentric orbit, filling its tidal lobe at pericenter. For HLX--1 the (assumed sub--Eddington) accretor must be an IMBH.
\citet{Lasota11} considered a model like this, but found that
to account for the observed outburst timescales it required a stellar orbit perilously close to becoming unbound
\citep{Godet14}, and very non-standard accretion disk structures \citep{Webb14}. Accordingly we retain (b) and consider instead the effect of dropping (c).

\section{Application to Super--Eddington Systems}
\label{sec:appsup}

The view that most if not all ULXs are super--Eddington systems has been greatly encouraged by recent observations giving accretor masses in the stellar range. This requires that the observed apparent X--ray luminosities exceed Eddington.
Most spectacular is the case of M82 X--2, which was discovered to be an accretion--powered X-ray pulsar with an apparent X-ray luminosity of $\sim 100 L_{\rm Edd}$ \citep{{Bachetti2012}}. Another ULX, the X-ray source P13 in the galaxy NGC 7793, was shown to have a $\la 15 \rm\, M_{\odot}$ black hole \citep{P13}. 
Finally, the observed properties of the source NGC 5907 ULX1 seem to be incompatible with the presence of an IMBH \citep{Walton14}. All these three ULX show flux variations by factors $\sim 100$, but the observational coverage is too sparse to determine the nature of the variability. 

Nevertheless one needs to consider the origin of transient behavior for these systems also. The solid observational evidence for super--Eddington systems makes it sensible to apply the disk instability model to such objects. Here we will show that the well-observed outbursts of HLX-1 can be explained by the TVDIM if it is a super-Eddington system.

Observers within the radiation beam of super--Eddington transients see them as ULXs. A solution like this is possible for HLX--1 \citep{KingLasota14} if we disregard the outburst behaviour. King \& Lasota suggested that the precession of the radiation beam might produce a similar light curve, and assumed a black hole mass $m \simeq 10$, determining an Eddington factor $\dot m \simeq 110$. 

But we see that combining (\ref{lsph}, \ref{eddratio2}) allows a self--consistent solution in which the X--ray light curve {\it is} given by disk instabilities instead of precession. This disk instability solution gives a lower accretor mass $m \simeq 3$ and a slightly higher Eddington factor $\dot m \simeq 170$. It requires a mean binary mass transfer rate $\sim 10^{-5}\msun\,{\rm yr}^{-1}$, which is typical for systems undergoing thermal--timescale mass transfer, like SS433 \citep{KTB00}. The companion star must be fairly cool to allow the the ionization instability. We note that the quasi--periodic recurrence of the outbursts would naturally appear if the outbursts are similar (as observed) and the disk is refilled at a roughly constant rate.
A system like this would have a period $\sim$ 50\,days.

If HLX--1 is subject to the thermal--viscous instability, only a stellar mass can account for the observed outburst timescales. These timescales are characteristic of such systems and no others.

\section{Conclusion}

We have shown that there is a simple relation between the accretor mass, and the peak outburst luminosity and decay timescale of X--ray outbursts. We used this to show that the observable X--ray population of elliptical galaxies probably consists largely of very long--lasting transients. We showed that the strongly--marked variations of the X--ray source HLX--1 cannot result from a disk instability if is the accretor is an intermediate--mass black hole. 

Acceptable solutions  for the disk instability picture are possible in two other ways. First, if the X--ray source is within the Local Group it may be bright sub--Eddington X-ray transient containing a stellar mass black hole. Alternatively if the source is in the galaxy ESO 243--49 it may be a strongly beamed stellar--mass system resembling SS433 undergoing outbursts driven by disk instabilities. We suggest that at least some of the variability of other, recently identified super--Eddington ULXs may result from the same process.

\bigskip
\bigskip
\section*{Acknowledgments}
We are very grateful to the anonymous referee for the very helpful criticism of the first version of this paper.
We thank Chris Done, John Cannizzo and  Chris Nixon for helpful conversations.
JPL acknowledges support from the French Space Agency CNES and Polish NCN grant UMO-2013/08/A/ST9/00795. 
Theoretical astrophysics in Leicester is supported by an STFC Consolidated Grant.
GD also acknowledges support from CNES.

\end{document}